# Controlled Electron Transport Through Single Molecules

Colin Lambert, Iain Grace, Theodoris Papadopoulos
Department of Physics,
Lancaster University, Lancaster, UK, LA1 4YB
Email: c.lambert@lancaster.ac.uk

*Abstract*— Using a first principles approach, we study the electron transport properties of a new class of molecular wires containing fluorenone units, whose features open up new possibilities for controlling transport through a single molecule. We show that the presence of side groups attached to these units leads to Fano resonances close to the Fermi energy. As a consequence electron transport through the molecule can be controlled either by chemically modifying the side group, or by changing the conformation of the side group. This sensitivity, opens up new possibilities for novel single-molecule sensors. We also show that transport can be controlled by tilting a molecule with respect to the electrode surfaces. Our results compare favorably with recent experiments.

*Keywords-- molecular electronics; transport; Fano resonances; nanoscale manipulation*

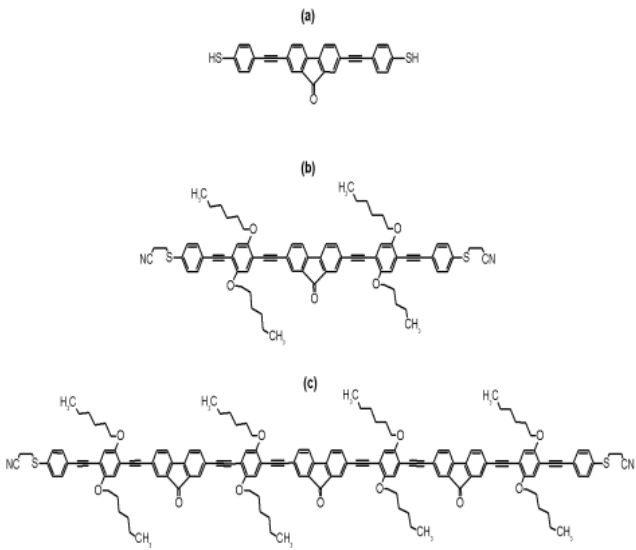

Figure 1. Three aryleneethynylene molecular wires of interest. (a) is approximately 2.5nm long, (b) approximately 4nm and (c) 7nm long.

## I. INTRODUCTION

Electric components built from single molecule are seen as a possible route to continued device miniaturization beyond 20nm feature sizes. The most common techniques for contacting single molecules to electrodes with nanometer-scale gaps involve STMs or break junctions. At present there exists no scalable technology capable of reliably contacting large numbers of single molecules. As a first step towards such a technology one can envisage first depositing contacts with a given spacing and then depositing molecules whose length matches the contact gap. Since the distance between lithographical grown contacts with reproducible spacings is typically 10nm or greater, rigid molecular wires with lengths of similar magnitude are needed. Recently a method has been developed [1-3] to synthesize rigid molecules with lengths up to 10nm comprising pi-conjugated oligomers based on rod like aryleneethynylene backbones which contain flourenone units. The presence of terminal protected thiol groups allows assembly onto gold surfaces and makes them ideal for use in single-molecule device fabrication Three of these molecules are shown in Fig. 1, and have lengths of 2.5, 4 and 7nm. The central part of all three molecules consists of a florenone unit (the 7nm contains three florenone units), which could be chemically modified e.g. by replacing the oxygen atom with a pyridine or bi-pyridine ring.

In this paper we compute electron transport properties of this family of different length wires. In particular we study how modification of the florenone units through the attachment of side groups can be used to control transport. In section II we develop the theoretical method employed in the calculations, which allows the transport properties of very long wires to be calculated and show results for transport through each of the molecules shown in Fig. 1.

In sections III and IV we investigate the effect of manipulating the molecule, first by changing the properties of groups attached to the florenone units and second by changing the angle between the long axis of the molecule and the gold electrodes. Recent experiments [4] using a STM technique have shown that increasing this tilt angle increases its conductance and in section IV we employ the method described in section II to explain this behavior.

## II. COMPUTATIONAL MODEL

The method employed uses a combination of the DFT code SIESTA [5] and a Green's function scattering approach, which was recently extended to yield the non-equilibrium molecular electronics SMEAGOL [6] code. To find the optimum geometry, the isolated molecules shown in Fig. 1 are relaxed using SIESTA until all force components on the atoms are smaller than 0.02 eV/Å. To ease the simulations, the long side chains ($C_6H_{13}$) attached to backbone, whose sole purpose is to aid solubility are replaced by methyl groups ($CH_3$). This has previously been shown to have a negligible effect on the conductance [7-8]. The next step is to extend the molecule to include the surface layers of the gold leads. When dealing with very long molecules the number of gold atoms in the extended contact should be as small as possible so that computing overheads are minimized, but the number of layers of gold should be great enough so that charge transfer effects at the gold molecule interface are included.

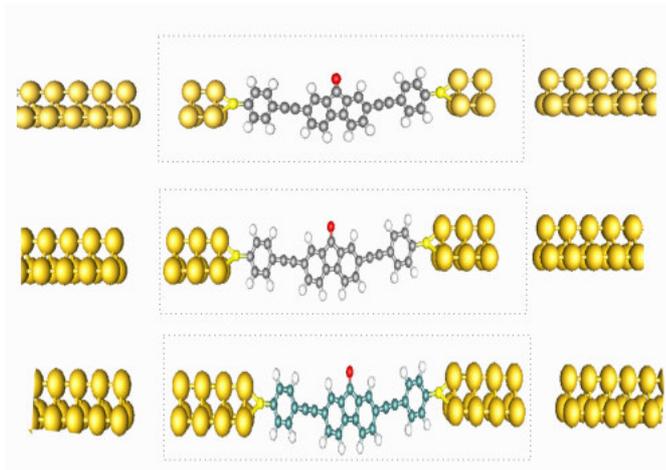

Figure 2. Three model structures of the extended molecule contacted between semi-infinite leads, with 2, 3 and 4 triangular layers of gold teated self-consistently.

The strongest bonding position for the sulfur atom on a gold (111) surface is in the hollow of three gold atoms. Therefore, to begin with we choose a very simple lead structure which consists of a triangular cross section, and then iteratively increase the number of layers that are included in the extended region (Fig. 2). Using a double-zeta basis plus polarization orbital basis (DZP), Troullier-Martins pseudopotentials [9] and the Ceperley-Alder LDA [10] to describe the exchange correlation the zero-bias, the underlying mean-field Hamiltonian of the extended molecule is computed, and combined via Dyson's equation with the Hamiltonian of the semi-infinite gold leads to yield the electron transmission coefficient T(E).

To illustrate some key calculational issues, we use the 2.5nm molecule shown in Fig. 1(a) as a test molecule and explore the outcome of varying certain parameters in the calculations. The effect of iteratively increasing the size of the contact region can be seen in Fig 3. Here we compute T(E) for electrons of energy E, for the 2.5nm molecule with the triangular lead cross section shown in Fig 2. A value of 2.1 Å was chosen as the gold sulphur bond length, since this is the calculated optimum length for a hollow site. Fig 3(a) shows the transmission through this molecule with 2, 3 and 4 layers of gold. These results show typical behaviour for transport through a molecule, with resonant peaks on either side of the Fermi energy (= 0eV) corresponding to the energy levels of the molecule. Fig. 3(a) shows that as the number of layers increases from 2 to 4, there is a noticeable difference between transmission through the HOMO states, whereas increasing the number of layers from 5 to 6 (shown in Fig. 3(b)) shows that the results have converged.

Using leads with a minimal cross-section significantly reduces the computational expense of the calculation. To investigate if this has any effect on the transport characteristics we increase the lead cross-section to a 3 by 3 cell, shown in the inset of Fig. 4, and recalculate the transmission coefficient for the 2.5nm molecule. The contact distance is kept the same and the bonding site is again the hollow between three gold atoms. Comparing this result (Fig. 4 solid line) with Fig. 3(b) the main difference between the curves occurs at E = -1.2eV. This reflects the presence of a band gap in the triangular leads which is absent from the larger cross-section wires. Since this difference is far from the Fermi energy it does not affect the low bias transport properties of interest in this paper.

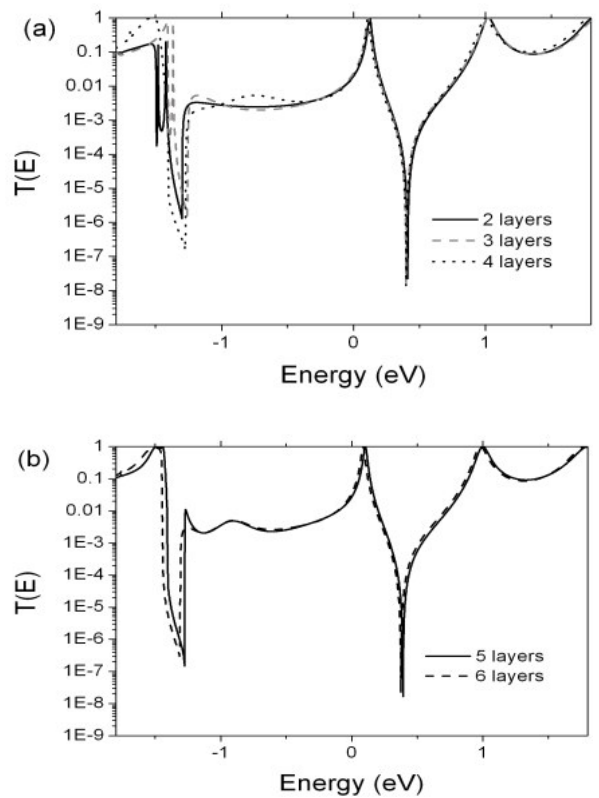

Figure 3. (a) Computed transmission coefficient as a function of energy for the 2.5 nm molecule of Fig. 1, with 2, 3 and 4 layers of triangular gold treated self consistently. (b) Transmission coefficients for the 2.5nm molecule with 5 and 6 layers. Comparison with Fig. 3 (a) shows that the transmission has converged.

So far we have used a DZP basis in all the calculations. In section IV, where we investigate the effect of tilting a molecule, large surface-area leads will be required to correctly include the bonding at a large tilt angle. Therefore in this case, to reduce the computational expense, the basis set will be reduced to a minimal single-zeta (SZ) basis. To see the effect of reducing the basis set, we re-compute the transmission through the 2.5nm molecule contacted to the 3 by 3 lead using a SZ basis. The results are plotted in Fig. 4 (dashed line) and show that the main change is a shift in the positions of the resonances, which is mainly due to a change in the position of the predicted Fermi energy. In addition, there is a reduction in the gaps between resonances. Despite these changes, the overall shape of the curve remains the same, which allows one to extract qualitative features, even with a minimal basis set.

We now examine the length dependence of these molecular wires by calculating the transmission coefficient for the 4nm and 7nm molecules shown in Fig. 1(b) and (c) respectively. In this calculation we use a DZP basis and simple triangular leads. Fig 5(a) (solid line) shows that the transmission through the 4nm molecule is very similar to that through the 2.5nm molecule (Fig. 3(b)) with the appearance of resonant peaks and a Fano resonance close to the Fermi energy. The main difference is a reduction in the off resonant transport and the appearance of an extra LUMO resonance.

For the 7nm molecule, which is composed of three flourenone units, T(E) is shown in Fig 5(b). There are now three Fano resonances close to the Fermi energy and the spacing between LUMO resonances has again decreased. The appearance of the three Fano resonances would suggest that the florenone units and their attachments are responsible for these resonant features. We explore the properties of these structures in more detail in section III, and show how they provide a useful route to controlling transport through a molecule.

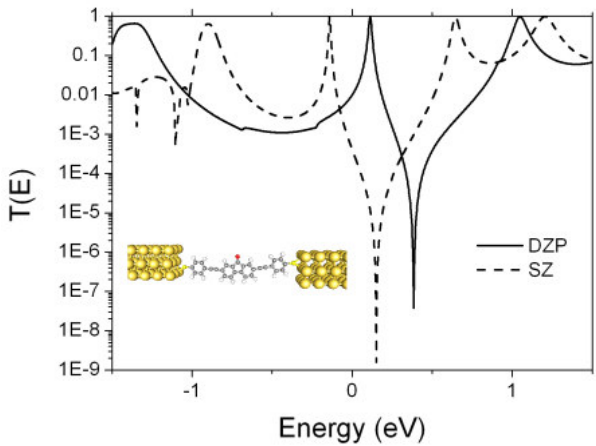

Figure 4. Computed transmission coefficient as a function of energy for the 2.5nm molecule, with a 3 by lead cross section (inset) for a DZP basis (solid) and a SZ basis (dashed).

So far we have only been interested in the zero bias transmission coefficient T(E), whereas to make comparisons with experiment the I-V characteristics must be computed by self consistently treating the change in charge on the molecule under finite bias. Fig 5(a) (dashed line) shows the transmission with a voltage of 1.5V across the 4nm molecule, this shows very little difference to the zero bias case (solid line) and therefore the associated electric field has little effect on the transmission at low bias. The inset of Fig. 5(a) shows the LUMO resonance at 0.05eV in more detail and reveals that the electric field causes a slight shift in the position of the resonance.

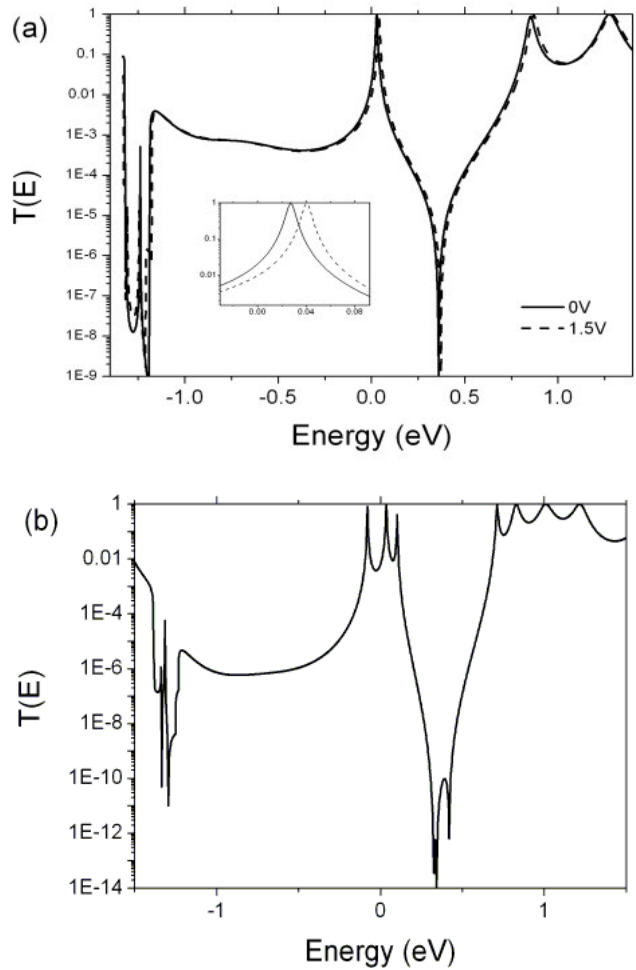

Figure 5. (a) Transmission coefficient for the 4nm molecule in Fig. 1(b). at zero bias (solid line) and with an electric field of 1.5V (dashed line). The inset shown zooms in on the resonance at 0.05eV. (b) Zero-bias transmission through the 7nm molecule shown in Fig 1(c)

The current through the molecule is computed using the Landauer formula

$$I = \frac{2e}{h} \int_{\mu_L}^{\mu_R} T(E, V_b) dE \qquad (1)$$

where $\mu_{L/R}$ are the chemical potentials in the left and right leads respectively and $T(E, V_b)$ is the transmission coefficient computed at a bias $V_b$. For the symmetric molecules of interest, $\mu_L = E_F - eV_b/2$ and $\mu_R = E_F + eV_b/2$, where $E_F$ is the Fermi energy, which is set to $E_F = 0$ in all figures. It is worth noting that equation (1) highlights a difficuly when comparing theoretical predictions for $T(E, V_b)$ with experimental I-V characteristics, because the computed current depends crucially on the position of $E_F$. Since the current implementation of SIESTA does not contain self-interaction corrections, the accuracy of the predicted $E_F$ is not known. Therefore when comparing with experiment, it may be necessary to treat $E_F$ as a free parameter to obtain a best fit.

For $E_F = 0$, Fig. 6 shows the calculated current through the 4nm (solid line) and 7nm molecules (dotted line). In both cases a current step appears at low voltages, due to the presence of transmission resonances close to $E_F$ and the magnitude at 1V is in the micro-amp range. Remarkably, the current through the 7nm molecule grows more quickly than the 4nm molecule, even though its off-resonance transmission is much smaller. This is due to the three Fano resonances close to the Fermi energy.

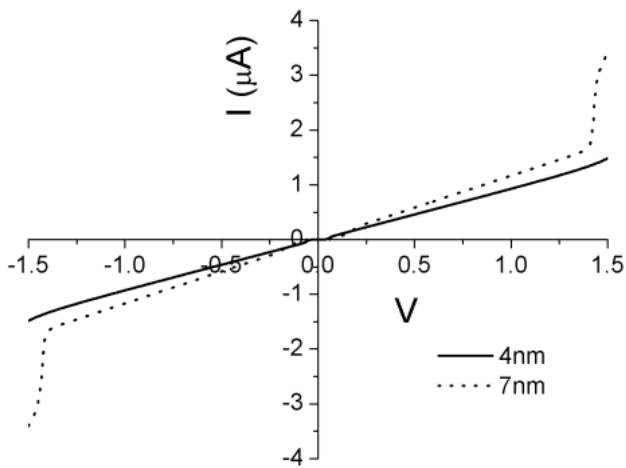

Figure 6. I-V behaviour for the 4nm (solid) and the 7nm (dotted) computing using eq. (1)

### III. FANO RESONANCES

The appearance of asymmetric Fano resonances [11], which comprise of a resonant peak followed by an anti-resonance, in the transmission coefficients of the three molecules studied so far means that a more complex picture than that suggested by a simple Breit-Wigner description of resonant transport [12] is needed. In this section we demonstrate that anti-resonances appear due to quasi-bound states associated with the side groups attached to the main axis of the molecular wire. In the results seen so far this would suggest that the oxygen atoms attached to the florenone units are responsible for these resonances. In the transmission coefficients of the 2.5nm and 4nm molecules (Fig. 3(b) and 5(a)), which contain one florenone unit there is one Fano resonance close to the Fermi energy. While in the 7nm molecule, which contains three florenone units there are three Fano resonances in close proximity at the Fermi energy.

To demonstrate that the Fano resonances are associated with oxygen atoms, we artificially set to zero the chemical bonds between the oxygen atom and the molecule backbone. Fig. 7(a) (dotted line) shows the effect of this trick on the 4nm molecule. This shows that the artificial removal of the bonds destroys the Fano resonance close to the Fermi level, while the rest of the transmission spectrum remains largely unaltered.

The ability to manipulate Fano resonances, by chemical means or otherwise, opens up new possibilities for controlling transport through single molecules. To explore this possibility in greater detail, the molecular wires in Fig. 1 have been synthesized with a pyridine and bi-pyridine molecule replacing the oxygen atom of the central florenone unit [2]. These new molecular wires open up the possibility of altering the properties of the attached side group for example through protonation and also by changing the conformation of the side group.

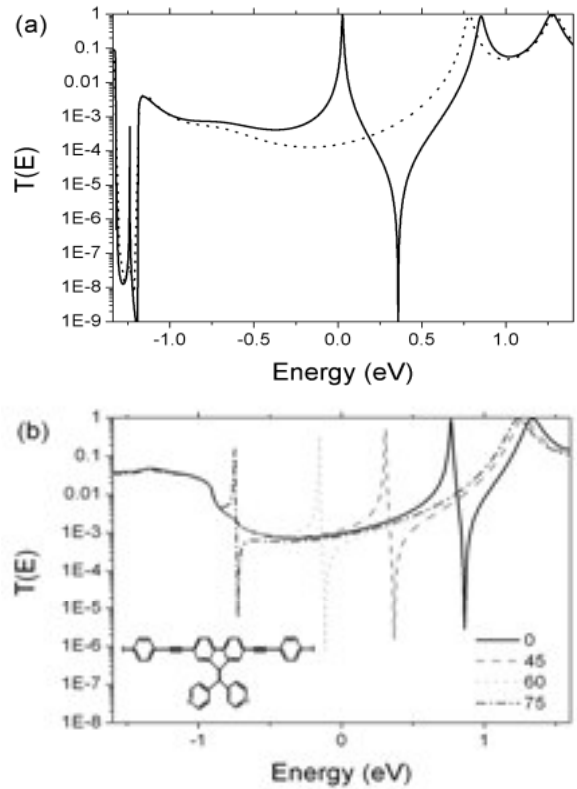

Figure 7. (a) Zero-bias transmission coefficient against energy for the molecule having an oxygen atom as a side group (solid line) and with the oxygen bond artificially removed (dashed line). (b) Transmission against energy for the bipyridine attached molecule (inset) for rotation angles from 0° to 90°.

To demonstrate that Fano resonances are a generic feature of molecular wires with attached side groups, we examine the molecule shown in the inset of Fig. 7(b). This is the 2.5nm molecule of Fig. 1(a) with the oxygen atom replaced by a bi-pyridine molecule. In section IV we model transport through a related molecule used in recent experiments, which control the conformation of the molecule in an electrical junction. Therefore, we model transport through the molecule shown in 7(b) for several conformations of the side group.

The transmission coefficient through this new 2.5nm molecule, obtained using a 3 by 3 lead structure and a DZP basis, is shown in Fig. 7(b) (solid line). This shows similar behavior to the transmission coefficient in Fig. 3(b), except the Fano resonance is shifted to 0.8eV, due to the difference in level structure of the bi-pyridine compared with oxygen. Fig. 7(b) also shows the transport properties for different values of the rotation angle θ of the bi-pyridine side group. We choose θ = 0° when the two rings lie parallel to the molecule axis and θ = 90° when the two rings lie perpendicular to the transport axis. The computed transmission is shown for four values of θ. This demonstrates that while the Breit-Wigner resonance at 1.25eV is almost unaffected by such a change, the position of the Fano resonance is very sensitive to rotations of the side group.

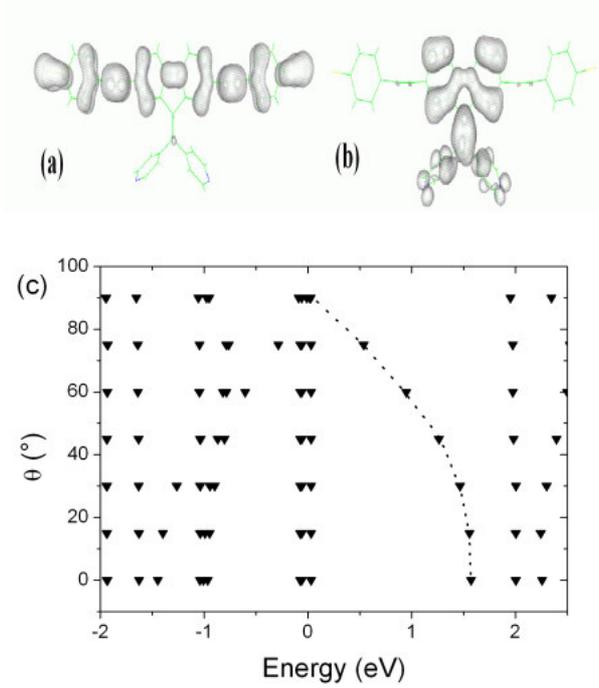

Figure 8. Surfaces of constant density of states for two energy levels of the isolated molecule, (a) E = 2.0 eV and (b) E = 1.56eV. (c) Energy levels of the isolated molecule as a function of rotation angle θ.

To understand this behavior and to demonstrate that the Fano resonance is associated with localized states on the side groups Fig. 8(c) shows the energy levels of the isolated molecule. Fig. 8(a) and 8(b) show plots of the local density of states (LDOS) for two energy levels of the isolated molecule; Fig. 8(a) shows the LDOS at an energy E = 2.0 eV, which clearly corresponds to states delocalized across the backbone with almost no weight on the bi-pyridine. This state is responsible for a Breit-Wigner resonance and as shown in Fig. 8(c), is almost independent of θ. On the other hand, for E = 1.56eV, Fig. 8(b) shows that the LDOS is localized on the flourenone unit. Fig. 8(c) shows that such states are sensitive to a change in θ, while most of the other levels remain unaffected. For example the level at E=1.56eV for θ = 0° moves to 0eV for θ = 90°. This behavior explains the transport results in Fig. 7(b), where the rotation of the side group alters the strength of the coupling to the backbone and causes a shift of the energy levels belonging to the bi-pyridine.

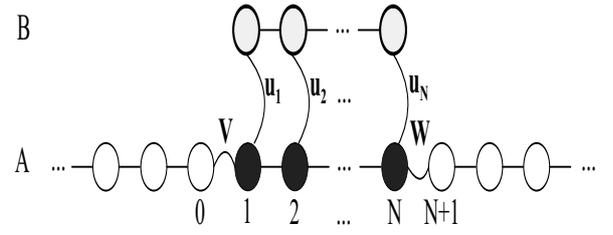

Figure 9. One Dimensional toy model of an infinite tight binding chain (A) which has a weakly coupled finite side chain (B)

We now develop a simple model describing the generic features of these resonances using the system sketched in Fig. 9 which consists of a backbone A composed of sites numbered i=1, 2,…,N, coupled to a side group B by matrix elements $H_1$. Sites i ≤ 0 belong to the left lead and sites i ≥ N+1 to the right lead. These are coupled to the ends of the backbone via weak hopping elements, V and W. The coupling matrix has the structure

$$H_1 = \begin{pmatrix} (H_1)_{k,1}^{A,A} & (H_1)_{k,1}^{A,B} \\ (H_1)_{k,1}^{B,A} & (H_1)_{k,1}^{B,B} \end{pmatrix} \quad (2)$$

and the Green's function G for the whole system is given by Dyson's equation $G = (g^{-1} - H_1)^{-1}$. For the purpose of computing transport properties we solve this expression for the the Green's function $G_{i,j}^A$ connecting sites (i,j) belonging to the backbone and/or leads only. This is obtained by expanding the right side of Dyson's equation to yield

$$G = g + gH_1g + gH_1gH_1g + gH_1gH_1gH_1g + \ldots \quad (3)$$

and noting that only the terms with even powers of $H_1$ contribute to $G_{i,j}^A$. Retaining only a single wavefunction $\Psi_n(k)$ of the side group (chain B) with eigenvalue ε, this yields

$$G_{i,j}^A = g_{i,j}^A + \sum_{k,l,m,p} \frac{g_{i,k}^A (H_1)_{kl} \Psi_n(l) \Psi_n^*(m)(H_1^*)_{mp} g_{p,j}^A}{E - \varepsilon - \Sigma} \quad (4)$$

where E is the energy of the propagating electrons and

$$\Sigma = \sum_{k,l,m,p} \Psi_n^*(k)(H_1)_{kl} g_{lm}^A (H_1^*)_{mp} \Psi_n(p) \quad (5)$$

In the absence of coupling to the side chain we assume that transmission through the backbone takes place through a single backbone state $\Phi_m(k)$, with a resonant energy $\varepsilon_0$. In this case, Σ simplifies to

$$\Sigma = \frac{\varpi \varpi^*}{E - \varepsilon_0 + i\Gamma} \quad (6)$$

where $\varpi = \langle \Psi | H_1 | \Phi \rangle$. In this expression, $\Gamma = \Gamma_1 + \Gamma_2$ is the broadening due to the coupling of the backbone to the leads, with $\Gamma_1 = |V\Phi_m(1)|^2 N_0(E)$, $\Gamma_2 = |W\Phi_m(N)|^2 N_{N+1}(E)$ and $N_0(E)$ ($N_{N+1}(E)$) is the local density of states for the left (right) contacts. This yields for the transmission coefficient $T = |\hbar v G^A_{N+1,0}|^2$,

$$T = \frac{4\Gamma_1\Gamma_2}{\left(E - \varepsilon_0 - \frac{\omega\omega^*}{E - \varepsilon}\right)^2 + \Gamma^2} \quad (7)$$

Equation (7) shows that when $\omega = 0$, no anti-resonance occurs and the transmission coefficient exhibits a single Breit-Wigner peak. More generally, equation (7) shows that transmission is a maximum at energies $E=\varepsilon_\pm$, where $\varepsilon_\pm$ are the roots of the equation $(E - \varepsilon_0)(E - \varepsilon) - \omega\omega^* = 0$ and vanishes when $E = \varepsilon$. For the case of identical leads and a symmetric backbone, eq (7) becomes

$$T = \frac{\Gamma^2(E-\varepsilon)^2}{(E-\varepsilon_+)^2(E-\varepsilon_-)^2 + \Gamma^2(E-\varepsilon)^2} \quad (8)$$

For small $\omega$, a standard Breit Wigner peak of width $\Gamma$ occurs in the vicinity of $\varepsilon_+ \approx \varepsilon_0$. In addition a Fano peak occurs in the vicinity of $\varepsilon_- \approx \varepsilon$ with width $\Gamma\omega\omega^*/(\varepsilon_0 - \varepsilon)^2$.

In Fig. 10 we show two representative plots of eq. (8). The solid line is obtained for values of $\varepsilon_- = 1.0$ eV, $\varepsilon = 1.1$ eV and $\varepsilon_+ = 1.5$ eV, and the dashed line corresponds to values of $\varepsilon_- = 1.51$ eV, $\varepsilon = -1.5$ eV and $\varepsilon_+ = 1.4$ eV. Comparing this result with the ab-initio data shown in Fig. 7(b) shows that with an appropriate choice of parameters, this equation captures the essential features of Fano resonances in molecular wires.

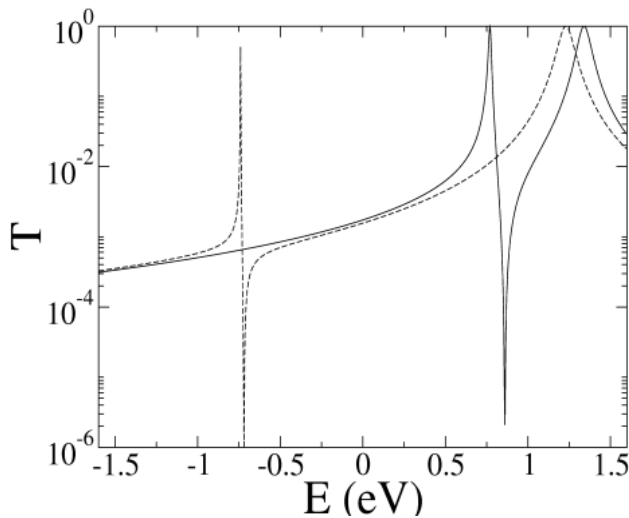

Figure 10. Numerical plot of eq. 7 when there is a shifting of the fano resonance

### IV. CONFORMATION-DEPENDENT TRANSPORT

Recently a new experimental method [4] has been developed using a STM technique which allows single molecules to be tilted, compressed or stretched within an electrical junction. The electrical behaviour of a conformationally-rigid molecular wire, shown in Fig. 11, was measured under such control and it was shown that the low bias conductance of the molecule increases very slowly at low tilt angles and then increases much more sharply for angles above 40°. We now apply the method described in section II to this problem to understand if this behavior is due to a stronger coupling of the molecule to the gold leads or a shorter transmission path.

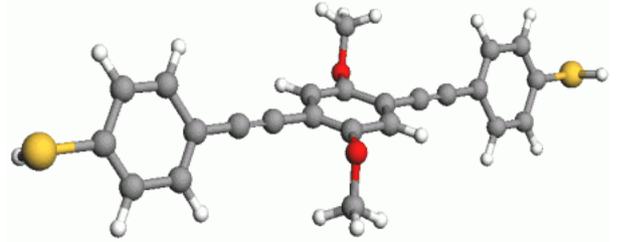

Figure 11. The 1,4-bis[4-(acetylsulfanyl)phenylethynyl]-2,6-dimethoxybenzene molecule.

Starting from the X-ray crystal structure [4], the relaxed geometry of the molecule was computed using SIESTA and as shown in Fig. 11, the central ring of this molecule is found to be rotated approximately 60° with respect to the plane of the two outer rings. The molecule was then extended to include the surface layers of the gold leads. To provide a large enough surface area to include the effects of creating a large tilt angle, each layer was chosen to have 15 gold atoms. In total, 5 such layers of gold were treated self-consistently on each contact.

In the STM measurements, precise details of the contact between the molecule and the electrodes are unknown and therefore we investigated the dependence of the conductance on tilt angle for a variety of docking sites. Overall, although the magnitude of the conductance is sensitive to the docking site, the dependence of conductance on tilt angle does not change appreciably. Therefore, in what follows, we only show results for the strongest bonding position, at both contacts, the sulphur atoms sit near the hollow between three gold atoms.

For all angles, the gold-sulfur distance has a fixed value of 2.1Å. The angle of tilt, $\theta$, is defined as the angle between the axis of the molecule and the normal to the gold surface, as shown in Fig. 12. One other important change that we investigate is the rigid rotation by an angle $\phi$ of the whole molecule with respect to the gold surface. As shown in Fig. 12, for $\phi = 0°$ and $\theta = 90°$ the end rings are perpendicular to the gold surface and for $\phi = 90°$ they are parallel to the surface.

For $\phi = 0°$, Fig. 13 shows the transmission coefficient $T(E)$ for electrons of energy $E$ passing through such a structure, for values of the tilt angle $\theta$ ranging from 10° to 70°. As $\theta$ increases the HOMO and LUMO resonances broaden and shift to lower energies, indicating that strength of the coupling between gold surfaces and the molecule increases with increasing $\theta$. For values of energy between 0.1eV and 0.6 eV, $T(E)$ increases with increasing angle, in agreement with experiment. In contrast, for values of $E$ between approximately 0.1 eV and –0.7 eV, $T(E)$ decreases with increasing angle.

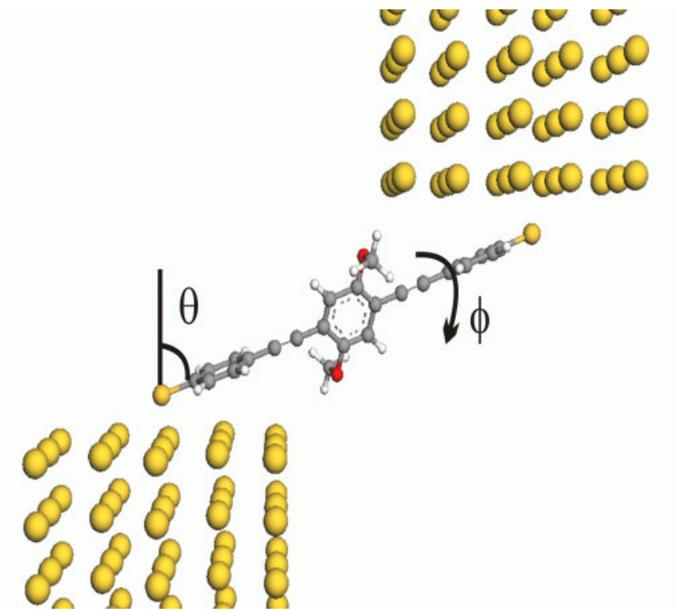

Figure 12. A cartoon showing the molecule between gold electrodes, with the tilt angle increasing as the spacing between the electrodes is decreased.

As noted in section II, the accuracy of the ab initio value for $E_F$ is not known and to obtain we have first adjusted $E_F$ to yield the best fit to measured I-V characteristics of the molecule. We found that a slight change from the ab initio value of $E_F = 0$ to $E_F = 0.2$ eV gave improved accuracy and therefore we adopt this value in what follows. Results for the zero-bias conductance $\sigma_M = (2e^2/h)T(E_F)$ as a function of tilt angle are shown in Fig 14. This shows that the magnitude of the predicted conductance is in good agreement with the experimental values. The main difference between the theoretical predictions and experiment data (hollow circles) is that if $\phi$ is fixed, the latter shows a steep increase in conductance for large tilt angles i.e. beyond $\theta \sim 35°$, which is not present in the theoretical results. We have also examined the effect of rigidly rotating the whole molecule about its axis by an angle $\phi$. This suggests that in the experiments, as $\theta$ is increased beyond $35°$ geometrical constraints cause $\phi$ to increase and therefore the experiment samples a succession of theoretical curves shown in Fig. 14.

In principle, the calculated dependence of $\sigma_M$ on $\phi$ could be a result of either changes in the hybridisation between the terminal sulphur atoms and the gold contact resulting in stronger electronic coupling from the contact into the molecular wire, or it could be a result of a change of the tunnelling path through the molecule. Fig. 15 (a) and (b) show the calculated density of states at the Fermi energy for $\phi = 20°$ and $\phi = 60°$ at a tilt angle of $\theta = 60°$. This demonstrates that the main source of the $\phi$-dependence is an increased hybridisation of the thiol and phenyl end groups with gold surface states.

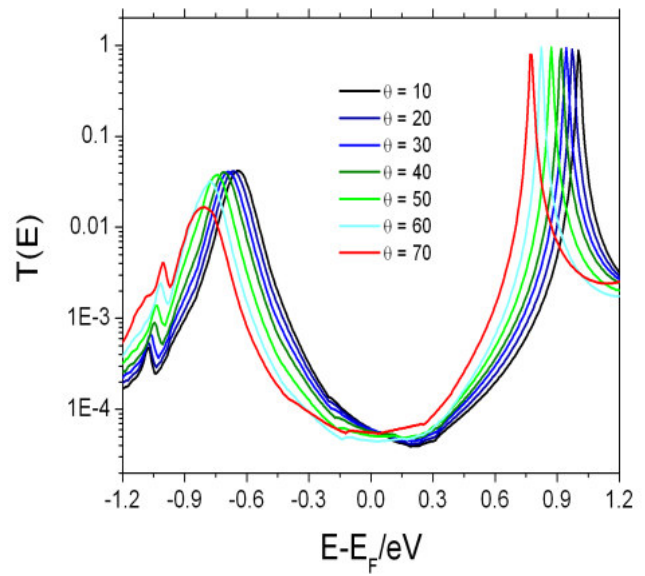

Figure 13. Calculated electron transport characteristics for the molecule contacted to Au(111) on three fold hollow sites, for tilt angles θ ranging from 10° to 70°.

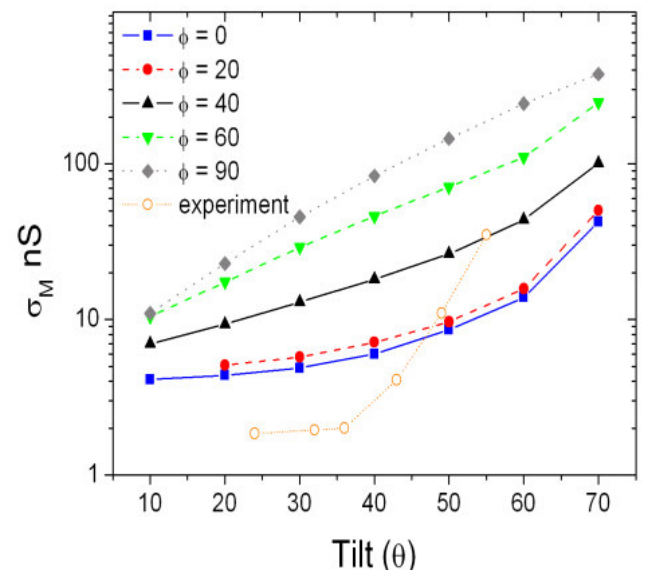

Figure 14. Conductance against tilt angle for five different rigid rotations of the molecule.

## V. SUMMARY

In this paper we have developed a theoretical method which allows the zero-bias conductance properties of long molecular wires to be calculated. We have shown that transport through this new class of molecular wires, composed of florenone subunits with side groups, is dominated by Fano resonances rather than Breit-Wigner resonances. One consequence is that the current through the 7nm molecule grows more quickly than the 4nm molecule. Another consequence is that electron transport through the molecule can be controlled either by chemically modifying the side group or by changing its conformation. This sensitivity which is not present in Breit Wigner resonances, opens up the possibilities of novel single-molecule sensors.

Since transport measurements on the molecules of Fig. 1 are not currently available, we have compared our techniques with experimental results for the tilt-angle dependence of electron transport through a related molecule. This suggests that experimental results can be reproduced theoretically, through a combined variation of angles θ and ϕ.

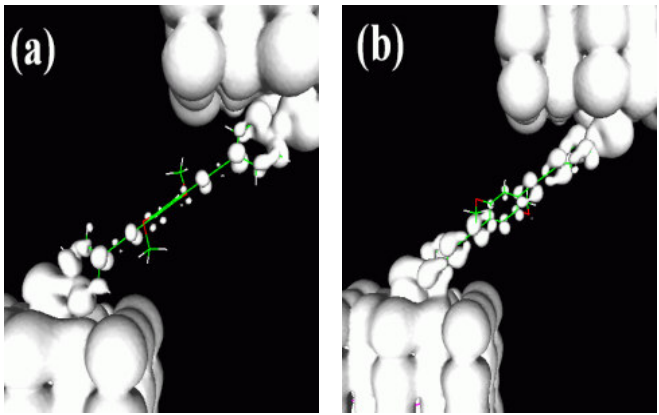

Figure 15. The local density of states at $E_F$ for a tilt angle θ = 60° for two rigid rotations of the molecule (a) ϕ = 20° and (b) ϕ = 60°

ACKNOWLEDGMENT

This work was supported by EPSRC under grant GR/S84064/01 (Controlled Electron Transport) (Durham and Lancaster), a Lancaster-EPSRC Portfolio Partnership and MCRTN Fundamentals of Nanoelectronics. It is a pleasure to acknowledge many fruitful discussions with M. Bryce, G. Ashwell, J. Jefferson, R. Nichols and W. Haiss.